\documentclass[preprint,showpacs,preprintnumbers,amsmath,amssymb]{revtex4}

\usepackage{graphicx}
\usepackage{dcolumn}
\usepackage{bm}

\begin{document}

\preprint{PRESAT7801}

\title{Magnetic orderings in Al nanowires suspended between electrodes}

\author{Tomoya Ono}
\affiliation{Research Center for Ultra-Precision Science and Technology, Osaka University, Suita, Osaka 565-0871, Japan}

\author{Shigeru Tsukamoto}
\affiliation{National Institute of Material Science (NIMS),\\1-2-1 Sengen, Tsukuba, Ibaraki 305-0047, Japan}

\author{Kikuji Hirose}
\affiliation{
Department of Precision Science and Technology, Osaka University, Suita, Osaka 565-0871, Japan}

\date{\today}

\begin{abstract}
A theoretical analysis of a relation between atomic and spin-electronic structures for the ground state of single-row aluminum nanowires suspended between Al(001) electrodes is demonstrated using first-principles molecular-dynamics simulations. We obtain a unusual result that a 3-aluminum-atom nanowire sandwiched between the electrodes does not manifest magnetic ordering although an isolated aluminum trimer molecule in a straight line is spin-polarized. On the other hand, a 5-atom nanowire exhibits ferromagnetic ordering, where three central atoms form a spin-polarized trimer. Moreover, in the case of an 8-atom nanowire, the middle atoms in the nanowire form two spin-polarized trimers with antiferromagnetic ordering.
\end{abstract}

\maketitle
As the miniaturization of electronic devices progresses, the fabrication of atomic-scale structures and the elucidation of their electronic properties have attracted much attention. In particular, the mechanical and electrical behavior of nanowires with a few atoms width and nanometer length has attracted a great deal of interest, because the properties of such minute systems with low dimensionality can be different from those of the bulk. First-principles molecular-dynamics (FPMD) simulations have been employed to investigate atomic and electronic properties of nanowires consisting of metallic elements. So far, many theoretical studies on single-row monatomic wires have been implemented \cite{okamoto,portal,torres,maria,ono-tsukamoto}; results of these studies generally showed good agreement in terms of atomic configurations, electronic structures and electron-transport properties of monatomic wires made of gold and sodium atoms. However, a serious discrepancy remains among theoretical studies on single-row aluminum wires: although aluminum is a nonmagnetic element, an {\it infinite} single-row wire in the ground state exhibits magnetic ordering due to spontaneous spin-polarization \cite{c,ono1,ono2}, whereas a single-row nanowire {\it suspended by aluminum electrodes} does not manifest magnetic orderings \cite{guo,nkobayashi,watanabe}.

On the experimental side, a large number of experiments concerning the generation of nanowires have been carried out using a scanning tunneling microscope and a mechanically controllable break junction \cite{datta-ruitenbeek}. Notable achievements are the recent fabrication of single-row nanowires of gold atoms and their observation by transmission electron microscopy, which reveals that the electric conductance is quantized in units of 2$e^2/h$ under their elongation and that the nanowires just before breaking form single-row chains with a conductance of 2$e^2/h$ \cite{takayanagi,kizuka,rubio,costa}. As for the nanowires made of aluminum atoms, although experiments have measured the conductance \cite{krans,scheer}, the stable single-row nanowires have not been visualized yet. There also remains the unanswered question regarding magnetic orderings in the single-row aluminum nanowires, and thus legitimate FPMD simulations are imperative for definite clarification of the magnetic properties of the nanowires.

In this paper, we explore the relationship between atomic and spin-electronic structures for the ground state of single-row aluminum nanowires suspended by Al(001) electrodes, based on FPMD simulations within the framework of the density functional theory. Our findings are that just as infinite single-row aluminum wires manifest magnetic ordering, the aluminum nanowires sandwiched between the electrodes exhibit a variety of magnetic properties (e.g., paramagnetic, ferromagnetic and antiferromagnetic orderings) depending on their length.

Our FPMD simulations are based on the real-space finite-difference method \cite{rsfd}, which enables us to determine the self-consistent electronic ground state and the optimized atomic geometry with a high degree of accuracy by means of the timesaving double-grid technique \cite{tsdg} and the direct minimization of the energy functional \cite{dmef}. The norm-conserving pseudopotential \cite{tmpp,kobayashi} is adopted and exchange-correlation effects are treated by the local-spin-density approximation \cite{lsda}. We employ a conventional technique that uses a three-dimensional periodic supercell and perform the Brillouin-zone integration only on the $\Gamma$ point. We take a cutoff energy of 25 Ry, which corresponds to a grid spacing of 0.63 a.u., and a higher cutoff energy of 225 Ry in the vicinity of nuclei with the augmentation of double-grid points \cite{tsdg}.

\begin{figure}[bt]
\includegraphics{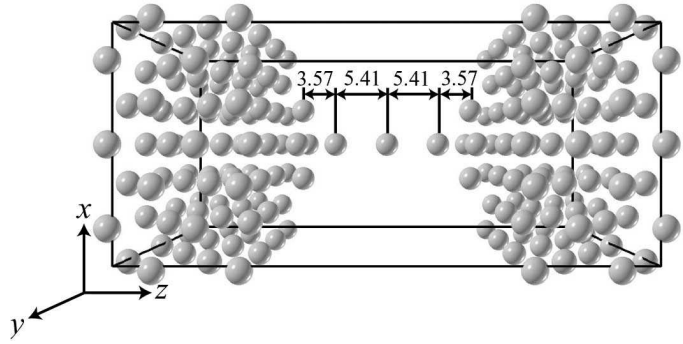}
\caption{Optimized atomic geometry of the 3-aluminum-atom nanowire suspended between Al(001) electrodes. Marked distances are values for projections of the distances between adjacent atoms onto the $z$ component, described in atomic units. The box indicates the supercell adopted in this calculation.}
\label{fig:1}
\end{figure}

Let us consider the nanowires composed of $N$ aluminum atoms ($N$=3, 5, and 8). The nanowire is connected to square bases at both ends, modeled after the [001] aluminum strands, and all of these components intervene between the electrodes produced from seven atomic layers of the Al(001) surface (see Fig.~\ref{fig:1} for illustration of the 3-atom nanowire). The distance between the atomic planes of the electrode surfaces and those of the bases is chosen to be 0.5$a_0$, where $a_0$ (=7.56 a.u.) is the lattice constant of the aluminum crystal. We adopt a supercell of size $3a_0 \times 3a_0 \times L_z$ which contains a total of $126+N$ atoms, where $3a_0$ is the lateral length of the supercell in the $x$ and $y$ directions perpendicular to the nanowire axis and $L_z$ is the length in the $z$ direction parallel to the nanowire. As an initial configuration of the atomic geometry, aluminum atoms in the nanowire are set at equal intervals of 5.20 a.u. in a straight line, where 5.20 a.u. is the critical interatomic distance for the breaking of the infinite single-row aluminum wire \cite{ono2}, and the distance between the end atoms in the nanowire and atomic planes of the bases is set to be 0.5$a_0$. We relieve the forces acting on the atoms of the nanowire until the maximum forces are smaller than 82.4 pN. During optimization, all of the atoms of the nanowire are treated dynamically, while the atoms of the electrodes and bases are kept frozen.

We first calculate the ground-state spin-electronic structure for the 3-aluminum-atom nanowire sandwiched between the aluminum electrodes. Although the density-functional calculation for an isolated aluminum trimer molecule in a straight line yields the result that the ground-state electron configuration is $^4 \Sigma_g$ \cite{comment1}, the energy bands for up- and down-spin electrons of this 3-atom nanowire spontaneously degenerate and only an ordinal nonmagnetic solution with no spin polarization is obtained. We show in Fig.~\ref{fig:1} the optimized bond lengths for 3-atom nanowire to explore why magnetic orderings do not emerge in the 3-atom nanowire; one can see that the two end atoms of the nanowire move toward the bases due to edge effects, and a trimer is not formed since the central atom of the nanowire is kept isolated. Thus, the spin polarization in the 3-atom nanowire is eliminated by the interaction between the electrodes and the end atoms in the nanowire.

\begin{figure*}[bt]
\includegraphics{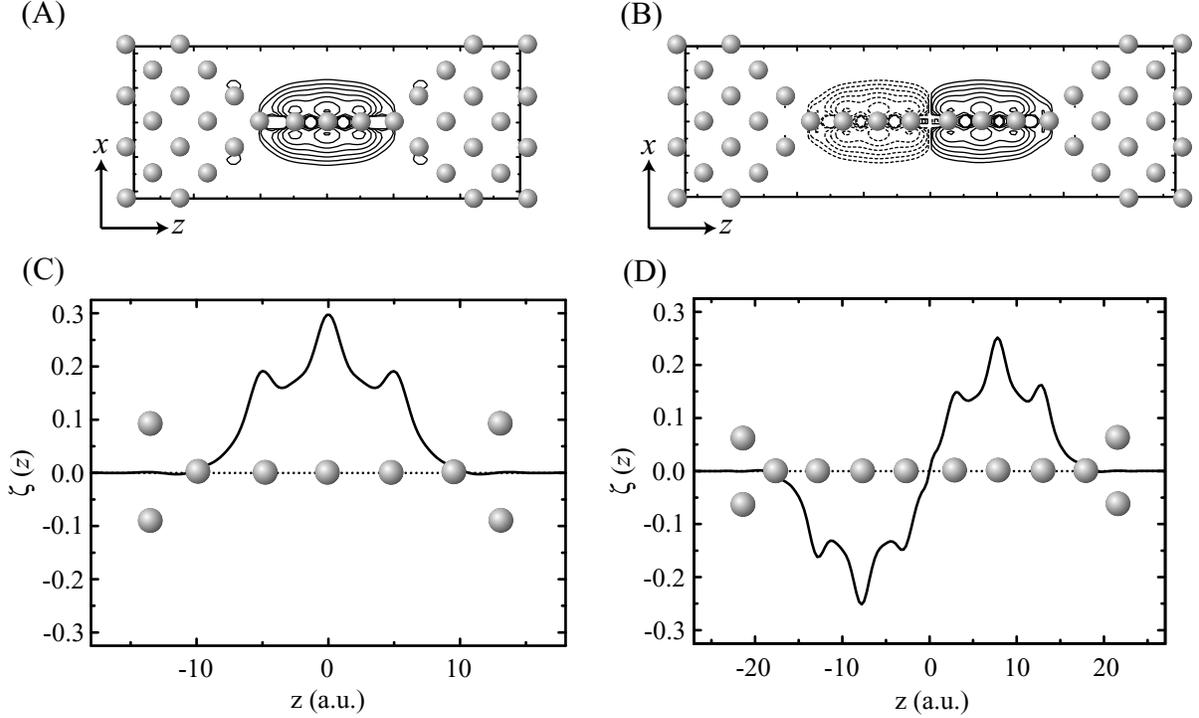}
\caption{Spatial distribution of the difference of the spin densities, $\rho_\uparrow (\mbox{\bf r})-\rho_\downarrow (\mbox{\bf r})$, for the ground state of the (A) 5- and (B) 8-aluminum-atom nanowires. Relative spin polarization $\zeta(z)$ for (C) 5- and (D) 8-atom nanowires plotted along the nanowire axis, where $\zeta(z) = n^-(z)/n^+(z)$ and $n^\pm(z)=\int \left[\rho_\uparrow ({\bf r}) \pm \rho_\downarrow ({\bf r}) \right] dxdy$. Gray spheres correspond to aluminum atoms. In (A) and (B), each contour represents twice or half density of the adjacent contour lines, and positive (negative) values of spin density are shown by solid (dashed) lines. The lowest contour represents 7.83 $\times$ 10$^{-5}$$e$/\AA.}
\label{fig:2}
\end{figure*}

Next, Fig.~\ref{fig:2} depicts the difference of the spin densities, $\rho_\uparrow (\mbox{\bf r})-\rho_\downarrow (\mbox{\bf r})$, and the relative spin polarization in lateral integration $\zeta(z)$, where $\zeta(z) = n^-(z)/n^+(z)$ and $n^\pm(z)=\int \left[\rho_\uparrow ({\bf r}) \pm \rho_\downarrow ({\bf r}) \right] dxdy$, for the ground state of 5- and 8-aluminum-atom nanowires. Spin polarization is clearly recognized and the nanowires are found to be in a nanoscale magnetic ordering in which the middle three atoms in the nanowire form spin-polarized trimers. In the case of the 5-atom nanowire, the total energy for the ferromagnetic ground state is lower than that of the nonmagnetic state by 52 meV per nanowire. In addition, we observe interesting antiferromagnetic ordering of the 8-atom nanowire suspended between the electrodes. The nanoscale antiferromagnetic ordering which originates from the two adjacent spin-polarized trimers formed inside the nanowire is observed. However, the total-energy differences per nanowire are 3 meV and 15 meV for nanowires in ferromagnetic ordering and paramagnetic ordering, respectively \cite{comment2}, and the energy difference between the 8-atom nanowires in antiferromagnetic and ferromagnetic states is so small that determination of the magnetic ordering is not feasible in the present scheme of the calculation.

We implemented a comparative analysis of the electronic and atomic structures of an infinite single-row aluminum nanowire in the previous study \cite{ono2}. An infinite aluminum nanowire in a magnetic state, where the atoms are uniformly spaced under stretching conditions, has a metallic character of a partially filled band that crosses the Fermi level. When it is elongated up to the average interatomic distance of 5.20 a.u., the spin density wave that arises from the Coulomb repulsion interaction yields spontaneous stabilization so that spin-polarized aluminum trimers with antiferromagnetic ordering emerge in the optimized infinite wire. Eventually, the trimerized structure with magnetic ordering observed in the 5- and 8-atom nanowires intervening between the electrodes is similar to that in the infinite wire.

In summary, we have examined the relationship between nanowire lengths and spin-electronic structures for the ground state of single-row aluminum nanowires connected to Al(001) electrodes. Our findings are probably the first theoretical indication that single-row aluminum nanowires suspended between the electrodes exhibit nanoscale magnetic orderings just as infinite single-row aluminum wires manifest magnetic ordering. We hope that our calculations open a possibility of a new type of magnetic materials, consisting solely of nonmagnetic elements, and encourage experiments toward to development of a new technology for spintronics in atomic nanowires.

This research was partially supported by the Ministry of Education, Culture, Sports, Science and Technology, Grant-in-Aid for Young Scientists (B), 14750022, 2002. The numerical calculation was carried out by the computer facilities at the Institute for Solid State Physics at the University of Tokyo, and Okazaki National Institute.

\newpage

\begin{references}
\bibitem{comment1}
The FPMD calculation shows that the optimized bond length of an isolated aluminum trimer in $^4 \Sigma_g$ state is 4.89 a.u.
\bibitem{comment2}
We confirmed that the 8-aluminum-atom nanowire with ferromagnetic ordering also forms trimerized structure in it.
\end{references}
\end{document}